\documentstyle[11pt,paspconf,epsf]{article}
\begin{document}

\title{Angular Momentum Transport in Simulations of Accretion Disks}
\author{James Rhys Murray}
\affil{Canadian Institute for Theoretical Astrophysics, McLennan Labs,
60 St George Street, Toronto M5S 3H8, Ontario, Canada}
\begin{abstract}
In this paper we briefly discuss the ways in which angular momentum transport
is included in simulations of non-self-gravitating accretion disks,
concentrating on  disks in close binaries.
Numerical approaches  fall in two basic categories; particle based Lagrangian
schemes, and grid based Eulerian techniques. Underlying the choice of
numerical technique are  assumptions that are made about disk
physics, in particular about the angular momentum transport mechanism.
Grid-based simulations have generally been of hot, relatively
inviscid disks whereas particle-based simulations are more commonly of
cool, viscous disks. Calculations of the latter type have been
instrumental in developing a model for the superhump phenomenon.
We  describe how we use an artificial viscosity term to introduce
angular momentum transport into our
smoothed particle hydrodynamics (SPH) disk code.
\end{abstract}
\section{Preamble}
The principal difficulty in modelling accretion disks stems from our
ignorance of the mechanism by which angular momentum is transported
through the disk. This process (or processes) must transfer
angular momentum from the inner disk to the outer disk, and also
convert kinetic energy to thermal energy in order to allow a mass flux
inwards through the disk. Our ignorance is usually reduced to a single
shear viscosity term in the Navier-Stokes equation, where the
kinematic shear viscosity 
\begin{equation}
\nu= \alpha c H.
\label{Shakura}
\end{equation}
In writing equation~\ref{Shakura}, Shakura \& Sunyaev (1973) assumed 
the unknown mechanism to be turbulent viscosity, in which case the
scale length of the largest turbulent eddies had to be less than the 
pressure scale height of the disk $H$, and the eddy velocity had to
be less than the sound speed $c$.
Thus the dimensionless parameter $\alpha$, that measures the
efficiency of the angular momentum transport, is expected to be less
than unity.
The accretion disks in outbursting dwarf novae for example, are
observed to have $\alpha \simeq
0.2$. The corresponding shear viscosity
is several orders of magnitude too large to be molecular. 

A requirement of our simulations then, is that they reproduce observed
rates of angular momentum transport. But does the transport occur via
disk scale structures such as nonlinear density waves,
that need to be resolved in the simulations, or
via very small scale, sub-grid, phenomena (e.g. turbulence)
that we can't explicitly include in the simulations but whose effects must be
allowed for in the equations of motion?  In the past, authors
investigating the former possibility have used high order, `inviscid',
grid based hydrodynamics algorithms, whilst the latter has been
investigated with `artificially viscous' particle schemes.

Sawada et al. (1986) simulated an accretion disk in a close binary
system using the second order Osher upwind scheme, a grid based
technique that does not require an artificial viscosity to resolve
shocks. In a two dimensional calculation set on the equatorial plane of
the binary,
they found that the secondary star launched spiral density waves at
the outer edge of the disk. These waves propagate inwards, carrying a
deficit of angular momentum which they transmit to the general flow
when they
damp. As a result there is a net outwards flux of angular momentum and
a mass flux inwards.
A self-similar analytic solution due to Spruit (1987) showed that a
disk in which spiral shocks provided the only mechanism for angular
momentum transport would have an effective $\alpha\la 0.01$. 
Calculations by Savonije et al. (1994) found that spiral
shocks were most efficient for hot disks with Mach numbers ${\cal M}=
v_\phi / c \le 10$. Here $v_\phi$ is the azimuthal velocity of the
gas. Disks in cataclysmic variables have Mach numbers closer to
30. Spiral shocks then are not the most important means for angular
momentum transport in these objects.  

From the above discussion we conclude that there is another mechanism
besides spiral shocks at work. At this stage the most likely
candidate is a weak field magnetohydrodynamic shear instability
discussed by Chandrasekhar (1960), but applied to the accretion disk
problem by Balbus \& Hawley (1991). Three dimensional simulations of
small regions of the disk (in the shearing sheet approximation) by
Stone et al. (1996) showed that magnetohydrodynamic turbulence
generated by the instability transported angular momentum with an
estimated efficiency $\alpha \la 0.01$, similar to the maximum
efficiency of spiral shocks. This is disturbing as it is an order of
magnitude less than observational estimates of  $\alpha$, which are obtained
by fitting axisymmetric viscous disk
solutions to the decay in luminosity  of dwarf novae  
after  outburst (Cannizzo, 1993).  
Dwarf novae in quiescence though, are thought to have an
effective $\alpha$ more in line with the Stone et al. result (although
in this case the observational results are less well constrained).

MHD turbulence is obviously a phenomenon that we cannot hope to
capture in a global disk simulation, however we can simply introduce
a shear viscosity term into
our numerical scheme to simulate the action of the weak field
instability.
Stone et al. suggest
that the  shear stress due to the instability 
is consistent with the Shakura \& Sunyaev
parametrisation (Equation 1).  

\section{Particle Methods and the Explanation for Superhump}
Lin \& Pringle (1976) developed a particle based scheme that included
an artificial method for `viscously' dissipating energy, in order to
study the evolution of disks in close binaries.
In their `sticky particle method', gas pressure is neglected and 
particles
move as test particles in the restricted three body problem.
 Then at the end of each time step a  grid is
placed over the computational domain.
Particle velocities are adjusted in order to minimise the kinetic
energy in each cell, whilst conserving linear and angular momentum.
Lucy (1977) cited the sticky particle method as being a a progenitor
of smoothed particle hydrodynamics. Having neglected pressure forces,
Lin \& Pringle excluded the possibility of 
angular momentum transport via density waves.
They simulated the evolution of accretion disks under steady
mass transfer from the companion star for three different binary mass
ratios, and in each obtained a steady state disk that was comparable
in size to the accreting star's Roche lobe.

Whitehurst (1988 a and b) developed a completely Lagrangian
scheme that included pressure forces, and subsequently found that
in close binaries with extreme mass ratios the secondary star is able
to excite a large eccentricity in the disk. This was particularly
exciting as some extreme mass ratio cataclysmic variables exhibited a
periodic luminosity variation known as superhump that was thought to
be the signature of an eccentric disk. The superhump period was always
a few percent larger than the orbital period. In Whitehurst's
simulation, the eccentric disk (as seen in the inertial frame)
exhibited a slow prograde precession. In his model the superhump
signal was due to the periodic stressing of the eccentric disk by the
tidal field, and the superhump period was the beat period between the
disk precession period and the binary period. 

Lubow (1991) showed analytically that eccentricity growth occurred
when the disk was large enough to encompass  
an eccentric inner Lindblad resonance. This could only happen when
the mass ratio $q=M_{\rm s}/M_{\rm p} \la 0.25$, where $M_{\rm p}$ is
the mass of the accreting star and $M_{\rm s}$ is the mass of the
companion.

Whitehurst's result was
verified and expanded upon by Hirose \& Osaki (1990) using the original
sticky particle method, and later by Murray (1996) using
SPH. The three particle techniques introduced viscous dissipation in
different ways, and the resulting kinematic viscosities had different forms.
In the sticky particle and SPH
simulations $\nu$ was constant, whereas Whitehurst's default choice
was $\nu \propto r^{-3/2}$ (Whitehurst also ran a simulation with  or
$\nu \propto r^{1/2}$). 
Thus, the detailed functional dependance of the viscous dissipation
has not proven to be critical. What is important and was common to  
the three sets of simulations was a
large  $\nu$  at large radii which allowed the disk to spread
out to the $3:1$ resonance. Heemskerk (1994) studied the phenomonon
using an Eulerian, grid-based scheme for solving  the inviscid
equations of hydrodynamics. Heemskerk performed some simulations using
only the $m=3$ component of the tidal field (that being the term
responsible for the Lindblad resonance), and found that the disk became
eccentric. When however he used the full tidal potential, the
accretion disk was unable to maintain contact with the resonance, even
though they had initially overlapped,  and
significant eccentricity growth did not occur. 
\section{Artificial viscosity in our SPH algorithm}
Smoothed particle hydrodynamics is a completely Lagrangian numerical
scheme, which uses an ensemble of particles to model a fluid (see
Monaghan 1992). The
fluid equations are replaced by a set of equations for the evolution
of the particles. For example, the fluid momentum  equation becomes, in
the SPH scheme, a set of equations for the forces on each particle
with spatial derivatives estimated by interpolating quantities from
neighbouring particles.
The `standard' SPH form of the momentum equation for particle $a$,
neglecting gravitational forces, is 
\begin{equation}
\frac{{\rm d} {\bf v}_a}{{\rm d} t} = - \sum_b m_b\,
(\frac{P_a}{\rho_a^2}+\frac{P_b}{\rho_b^2} + 
\frac{\beta {\mu_{ab}^2}-\zeta \bar c_{ab} \mu_{ab}}
{\bar \rho_{ab}}) \,\nabla_a
W({\bf r}_a-{\bf r}_b,h).
\label{eq:SPHmom}
\end{equation}
Here ${\bf r}_a$, ${\bf v}_a$ and $m_a$ are the position,
velocity and mass of particle $a$. $P_a$, $\rho_a$ and $c_a$ are the
pressure, density and sound speed of the fluid evaluated at 
${\bf r}_a$. $\rho$ is obtained by interpolation, and then $P$ and $c$
are obtained using the chosen equation of state.
The $\bar{X}_{ab}$ notation indicates the arithmetic mean
of quantity $X$ evaluated
at ${\bf r}_a$, and ${\bf r}_b$.
$\zeta$ and $\beta$ are the coefficients of the linear and
non-linear  artificial viscosity terms. $W({\bf r},h)$ is the
interpolation kernel.
\begin{equation}
\mu_{ab}= \frac{({\bf v}_a-{\bf v}_b)\cdot({\bf r}_a-{\bf r}_b)}
{({\bf r}_a-{\bf r}_b)^2+\eta^2},
\end{equation} 
with $\eta$ being a softening parameter usually set equal to one tenth the
smoothing length $h$. Artificial viscosity was originally
added to the SPH equations in order  to improve
the resolution of shocks, and is usually used with a switch that sets
it to zero when $({\bf v}_a-{\bf v}_b)\cdot({\bf r}_a-{\bf r}_b) >
0$. This ensures that dissipation only occurs for compressive flows. 
For disk simulations however, we want viscous dissipation to occur wherever
there is velocity shear so we use only the linear viscosity term with the
switch disabled. By letting the number of particles $n\to\infty$,
the smoothing length $h \to 0$, and approximating the summation
with an integral, we can obtain the continuum equivalent of the linear
artificial viscosity term. Following Pongracic (1988) and Meglicki et
al. (1993), we find that the `viscous' force per unit mass
\begin{equation}
{\bf a}_{\rm visc}=\frac{\zeta h \kappa}{2\rho}\,
({\bf \nabla}\cdot(c \rho {\bf S})+{\bf \nabla}
(c \rho {\bf \nabla} \cdot {\bf v})),
\label{eq:genvisc}
\end{equation}
where the deformation tensor  
\begin{equation}
S_{ab}=\frac{{\rm\partial}v_a}{{\rm\partial}x_b}+
\frac{{\rm\partial}v_b}{{\rm\partial}x_a}.
\end{equation}
$\kappa$, a
constant dependant only upon the smoothing kernel used, is $\frac{1}{4}$ for
the standard cubic spline kernel (in two dimensions).
If we assume the density and sound
speed vary on much longer length-scales than the velocity, we have
\begin{equation}
{\bf a}_{\rm v}=\frac{\zeta h c}{8}\,
(\nabla^2 {\bf v}+2{\bf \nabla}
({\bf \nabla} \cdot {\bf v})).
\end{equation}
In other words the linear artificial viscosity term generates both
shear and bulk viscosity  {\em in a fixed ratio}.
In the interior of the disk where ${\bf \nabla} \cdot {\bf v} \simeq
0$, we simply have a kinematic viscosity 
\begin{equation}
\nu=\frac{1}{8}\,\zeta c h.
\label{avec}
\end{equation}
In order to check the accuracy of 
equation~\ref{avec} we simulated the viscous
spreading of an axisymmetric disk with a  Gaussian initial 
surface density profile
\begin{equation}
\Sigma(r,t=0)=\Sigma_0\,e^{{\frac{(r-r_0)}{l^2}}^2}.
\label{Gauss}
\end{equation}

For this (two dimensional) calculation we laid down $22,400$ particles
in 21 equally
spaced concentric rings, varying the particle mass to generate the
Gaussian density profile. Gas pressure forces were switched
off, and a constant smoothing length was used.
We find that the actual shear viscosity generated by the
linear artificial viscosity term was within $\pm10\%$ of the value
 given by equation~\ref{avec}.
Figure 1 shows the close agreement between
the analytic solution for a Gaussian ring with viscosity given by
equation~\ref{avec}, and our SPH simulation. The test is described in
more detail in Maddison et al. (1996) along with a study of the
stability of a narrow ring of viscously interacting particles.

\begin{figure}
\plotone{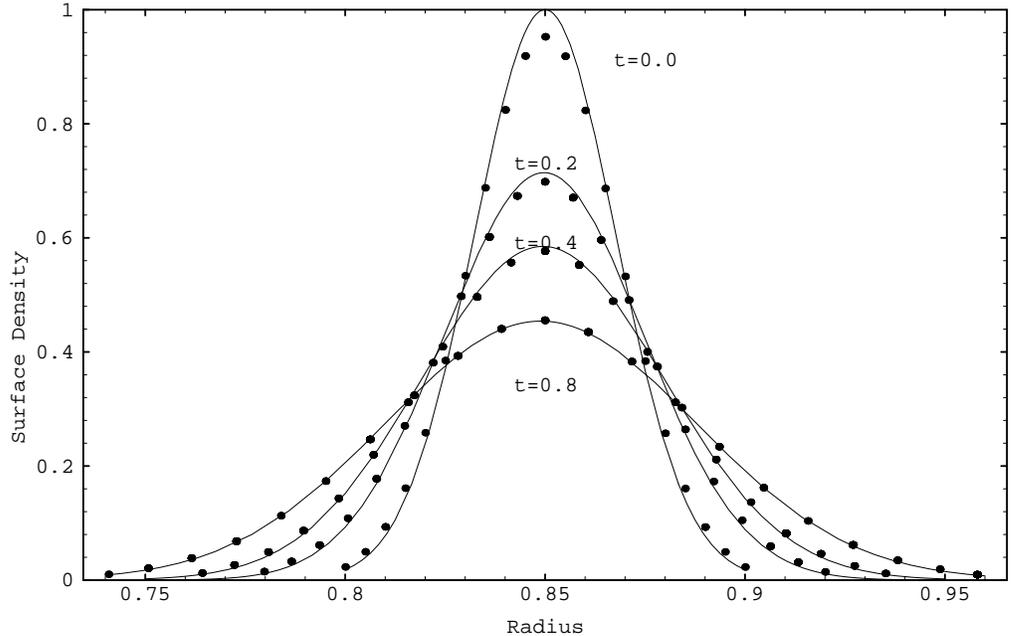}
\caption[ ]{Evolution of an axisymmetric ring with a Gaussian initial
density profile. The solid lines denote the analytic solution at the
times shown, and the heavy points show the SPH solution. In this
calculation we use $r_0=0.85$, $l=0.025$, $\Sigma_0=1.0$ and
$\nu=2.5\times10^{-4}$ (see
equation~\ref{Gauss}). Units have been scaled so ${\cal G} M=1$,
where $M$ is the mass of the central star. We used a fixed
 smoothing length $h=0.01$.}
\end{figure}

It is possible to derive SPH interpolant estimates of 
$\nabla^2\,{\bf v}$, and thus to directly include shear viscosity in the
SPH equations. Flebbe et al. (1994) and  Watkins et
al. (1996) introduced formulations and tested them with ring spreading calculations
similar to the one above, obtaining reasonable agreement between predicted
and actual values for $\nu$. However, their results are clearly more
noisy than those shown in figure 1. Both Flebbe et al., and Watkins et
al. must interpolate twice in order to obtain $\nabla^2\,{\bf v}$
which may increase the susceptibility of the calculations to the
single ring instability discussed in Maddison et al..
Most importantly, in the  Flebbe et al., and Watkins et al. formulations
the forces between two particles are not antisymmetric and along
the particles' line of centers. Therefore they do not exactly conserve
linear and angular momentum. With our term, linear and angular
momentum is conserved to machine accuracy. 

\section{Conclusions}
We  described above that  superhump simulations completed with a
low viscosity code gave results that were qualitatively different from more
viscous calculations. Clearly, we must quantify the effective shear
viscosity of accretion disk simulations if we are to make meaningful comparisons
with observations. The ring-spreading calculation is a simple way to
do this.
There are several ways to introduce a known shear viscosity into SPH
disk calculations. We described in detail a method which conserves linear and
angular momentum . Momentum conservation is important in disk
calculations that may run for thousands of dynamical time scales. We showed that
the effective $\nu$ introduced by our term could be accurately estimated with
equation~\ref{avec}.

\end{document}